\documentclass[11pt,fleqn]{article}
\usepackage{epsfig}
\usepackage{amsfonts}
\usepackage{times}
\usepackage{mathptm}
\newcommand{\Section}[1]
{\section{#1}\setcounter{equation}{0}}
\font\titlefnt=cmbx10 scaled \magstep2

\begin{document}
\mathindent 0mm
\newpage\thispagestyle{empty} 
\begin{flushright} HD--TVP--97--08\end{flushright}
\vspace*{2cm}
\begin{center} 
{\titlefnt Brownian motion in fluctuating periodic potentials\\ \vspace*{0.4cm}
}
\vskip2cm
Enrique Abad\footnote[1]
{E--mail: abad@tphys.uni-heidelberg.de}\\ 
Andreas Mielke\footnote[2]
{E--mail: mielke@tphys.uni-heidelberg.de}\\ 
\vspace*{0.2cm}
Institut f\"ur Theoretische Physik,\\
Ruprecht--Karls--Universit\"at,\\
Philosophenweg 19, \\
D-69120~Heidelberg, F.R.~Germany
\\
\vspace*{0.5cm}
\today
\\[0.5cm]
Accepted for publication in Ann. Physik (Leipzig).
\vspace*{0.7cm}
\noindent

{\bf Abstract}
\end{center}
This work deals with the overdamped motion of
a particle in a fluctuating one--dimensional periodic potential. If the
potential has no inversion symmetry and its fluctuations are
asymmetric and correlated in time, a net flow can be generated at
finite temperatures. We present results for the stationary
current for the case of a piecewise
linear potential, especially for potentials being close to
the case with inversion symmetry. 
The aim is to study the stationary current as a function
of the potential. Depending on the form of
the potential, the current changes sign once or even
twice as a function of the correlation time of the potential
fluctuations. To explain these current reversals, several mechanisms
are proposed. Finally, we discuss to what extent the model is
useful to understand the motion of biomolecular motors. 
\vspace*{0.2cm}\noindent

\vspace*{0.1cm}
PACS-Numbers: 05.40.+j, 05.60.+w, 87.10+e

\vspace*{0.2cm}

\vspace*{0.3cm}
\newpage
\topskip 0cm

\Section{Introduction}
Protein motors play an important role in intracellular transport
phenomena. In the cytoplasm of eucaryotic cells, certain
macromolecules move along a complex network of periodic polymer tracks
and transport organelles or vesicles containing chemicals.  These
ATP-hydrolysing macromolecules successively attach to and detach from
the biopolymer while walking on it. Structural analysis of several
motor proteins together with {\it in vitro} experiments have revealed
that they undergo a cyclic sequence of conformational changes to
convert the chemical energy from the ATP-hydrolysis into a
unidirectional movement along the biopolymer.  (see \cite{Svoboda} and
the references therein).  Although this mechanochemical conversion
mechanism is not yet fully understood, there is no macroscopic thermal
or chemical gradient in the cell medium that determines the direction
of movement. Thus an interesting question arises: which properties of
the system {\it protein+track} favour one direction of movement rather
than the opposite?  This is certainly a difficult matter: one knows
motor proteins of the same family that move in different directions
along the same track.

Recently, some progress has been made in attempting to answer the
above question. Apparently, the biasing of the catalytic cycle of the
ATP hydrolysis is due to time-correlated chemical fluctuations induced
by far-from-equilibrium concentrations of the reactants.  To describe
the motion of molecular motors, Magnasco considered the model of a
Brownian particle moving in a one-dimensional ratchet-like potential
driven by a symmetric stochastic force \cite{Magnasco}.  A similar
model has been used by Feynman {\it et al.} \cite{Feynman} to
illustrate the Second Law of Thermodynamics:  Thermal fluctuations
cannot produce a transport. (The example is originally due to 
Smoluchowski \cite{Smoluchowski}.) But if the fluctuations of the force are
time-correlated, the restrictions of the Second Law of Thermodynamics
cease to apply and the particle shows a net movement that can even
overcome the action of an external force: The asymmetric ratchet works
as a mechanical rectifier.  Both ingredients, the time correlation of
the symmetric fluctuations and the lack of inversion symmetry of the
potential yield a non vanishing current.
    
Astumian and Bier \cite{Ast1} proposed a different picture for the
motion of motor proteins. They studied the problem of a strongly
damped particle moving in a ratchet-like potential that fluctuates
dichotomously, i.e. between two different states \cite{Ast1}. The
two different states of the potential model the electrostatic
interaction of the motor protein with the periodic charge pattern of
the biopolymer. The charge of the protein is changed when ATP binds to
it.  As a result, the potential changes as well. In this model, a net
current is obtained at finite temperatures if the coloured potential
fluctuations are asymmetric.  Similar models have also been proposed
by Prost {\it et al.}  \cite{Prost,Chauwin}. These authors argued that the
two states correspond to the attached and the detached state of the
motor protein. For a recent overview on this and related models we
refer to \cite{AstR}, where the reader can find a large list of
relevant literature on the subject.

Although a simple dichotomous process contains some essential features
of the movement of motor proteins, it can on no account provide a
realistic description of the ATP hydrolysis. The main problem is that,
for a ratchet-like potential, the direction of the current is fixed.
It is therefore interesting to study other noise processes as well.
Mielke recently developed a method that applies both to Magnasco's
model of a fluctuating force \cite{Mi1} and to the model by Astumian
and Bier \cite{Mi2}. For a fluctuating sawtooth potential, he observed
in several cases a current reversal for slightly different parameter
sets of the multiplicative noise. Despite its simplicity, the model
shows that small modifications of the motor protein suffice to make
it move in the opposite direction.  Other noise processes that can
take not only two values as the dichotomous process can also be
motivated from the biological situation of the motor protein.
Typically, the motor protein undergoes several conformational changes,
so that the interaction between the motor protein and the substrate
must be described by several interaction potentials, not only two.

In this paper, we apply the method developed by Mielke to treat the
movement in a fluctuating potential.  Up to now, most efforts have
concentrated on studying the influence of the noise parameters on the
induced stationary current. Therefore calculations were performed
taking the simplest asymmetric potential, a sawtooth ratchet. We
rather focus on how the current varies when the geometry of the
potential is changed. This question is of general interest, since one
might expect new phenomena in a more complex potential. It is also of
relevance for the question, whether or not these models can be used to
construct a realistic picture of motor proteins. As we will show, the 
stationary current, which is the most important quantity to be calculated,
depends essentially on the specific form of the potential. Therefore
we conclude that a quantitative comparison between experimental findings
and theoretical predictions from the simple model should be taken with
care. A small change of the potential may change the picture quantitatively
and even qualitatively.

After defining the model, we derive a set of tridiagonal recursion
relations to calculate the current and the stationary probability
distribution. In Section 3 we take these relations as a starting point
to discuss the limiting case of white noise and compute the current up
to second order in the correlation time. In Section 4 we present exact
numerical results for piecewise linear potentials. It turns
out that direction and rate of the movement depend strongly on the
details of the potential.  Section 5 summarizes the main conclusions
of the present work and discusses the relevance of our results for the
motion of molecular motors.
\Section{Definition of the model}
The overdamped Brownian motion of a particle in a fluctuating periodic
potential can be described by a Langevin equation,
\begin{equation}
  \label{lan}
  \frac{dx}{dt}=-z(t)\frac{\partial V(x)}{\partial x}+\sqrt{2T}\xi(t).
\end{equation}
We have chosen the units so that the friction coefficient is
unity. The second term on the right hand side describes the
thermal noise, as usual one has
\begin{equation}
  \label{xi_corr}
  \left\langle \xi(t)\right\rangle=0 
  \quad 
  \left\langle \xi(t)\xi(t') \right\rangle=\delta(t-t').
\end{equation}
The first term describes a fluctuating periodic potential.
$V(x)$ is periodic with period $L$. 
We restrict ourselves to the case where the amplitude $z(t)$
of the potential $V(x)$ fluctuates. This case has
also been studied by Astumian and Bier \cite{Ast1}
and by Prost {\it et al.} \cite{Prost}. As mentioned above, they
discussed only the case where $z(t)$ takes two different
values. We consider the more general case where $z(t)$ is
a Markov process described by a general Fokker--Planck equation
\begin{equation}
  \label{fpz}
  \frac{\partial p(z,t)}{\partial t}=M_zp(z,t).
\end{equation}
The operator $M_z$ has one eigenvalue $\lambda_0=0$.
The corresponding right eigenfunction $\phi_0(z)$ is the
stationary distribution of the Markov process $z(t)$.  The other
eigenvalues are non--positive. The eigenvalue equation of
$M_z$ is 
\begin{equation}
  \label{Mz}
  M_z\phi_n(z)=-\lambda_n\phi_n(z),
\end{equation}
where $\phi_n(z)$ are the right eigenfunctions of $M_z$.
We assume that $0<\lambda_1\le\lambda_2\le\dots$. 
Without loss of generality we take $\langle z\rangle>0$.
The time-dependent autocorrelation function of $z(t)$ satisfies
\begin{equation}
  \label{z_korr}
  \left\langle z(t)z(t') \right\rangle-\langle z\rangle^2 
  \propto e^{ -|t-t'|/\tau}
\end{equation}
for large $|t-t'|$.
The correlation time $\tau$ is determined by the largest negative
eigenvalue of $M_z$, i.e.
$\tau=\lambda_1^{-1}$.
We restrict ourselves to a class of processes for which
\begin{equation}
  \label{rekur_phi}
  z\phi_n(z)=\gamma_{n,n+1}\phi_{n+1}(z)+\gamma_{n,n}\phi_n(z)
  +\gamma_{n,n-1}\phi_{n-1}(z),\qquad n\ge 1.
\end{equation}
We assume that the stationary distribution $\phi_0(z)$ of $z$
is normalized to unity. 
Due to the recursion relations (\ref{rekur_phi}) the eigenfunctions 
$\phi_n(z)$ can be written as $\phi_n(z)=g_n(z)\phi_0(z)$
where $g_n(z)$ are orthogonal polynomials with respect to the
weight function $\phi_0(z)$,
\begin{equation}
  \int dz g_n(z)g_m(z)\phi_0(z)=\delta_{n,m}.
\end{equation}
This class of Markov processes is very general. It contains
many processes that occur in typical situations such as
the Ornstein--Uhlenbeck process, the dichotomous process,
sums of dichotomous processes, and kangaroo processes.

Having defined the stochastic process $z(t)$ via the Fokker--Planck
equation (\ref{fpz}) it is natural to work with a Fokker--Planck
equation for the problem described by (\ref{lan}). To do this
one has to introduce a joint probability density $\rho(x,z,t)$
for $x=x(t)$ and $z=z(t)$. The Fokker--Planck equation for
$\rho(x,z,t)$ can be written in the form
\begin{equation}
  \label{fp}
  \frac{\partial \rho(x,z,t)}{\partial t}=
  -\frac{\partial}{\partial x}
  \left(zf(x)-T\frac{\partial}{\partial x}\right)\rho(x,z,t)
  +M_z\rho(x,z,t).
\end{equation}
In the following we will only consider static properties
of our model. Therefore it is sufficient to calculate the
static joint probability density $\rho(x,z)$. It satisfies
(\ref{fp}) with vanishing left hand side. For the
class of processes under consideration, it is suitable
to expand $\rho(x,z)$ in the complete system of 
right eigenfunctions $\phi_n(z)$ of $M_z$,
\begin{equation}
  \label{rho_stat}
  \rho(x,z)=p_0(x)\phi_0(z)+\sum_{n=1}^\infty(-1)^n\phi_n(z)p_n'(x).
\end{equation}
Using the equations (\ref{rekur_phi}) one derives
easily a set of recursion relations for $p_n(x)$.
\begin{equation} 
  \label{rekur_p0}
  J=\gamma_{0,0}f(x)p_0(x)-Tp'_0(x)-\gamma_{0,1}f(x)p_1'(x),
\end{equation}
\begin{equation} 
  \label{rekur_p1}
  \gamma_{0,1}f(x)p_0(x)=
  \lambda_1p_1(x)+\gamma_{1,1}f(x)p_1'(x)-Tp_1''(x)-\gamma_{1,2}f(x)p_2'(x),
\end{equation}
\begin{equation} 
  \label{rekur_pn}
  \gamma_{n-1,n}f(x)p_{n-1}'(x)=
  \lambda_np_n(x)+\gamma_{n,n}f(x)p_n'(x)-Tp_n''(x)
  -\gamma_{n,n+1}f(x)p_{n+1}'(x)\mbox{ for }n>1.
\end{equation}
$J$ is an integration constant which has a simple physical meaning,
it is the static current.
A slightly different 
derivation of these equations has been given in \cite{Mi1,Mi2}. In
general it is not possible to solve these equations, but they provide
a good starting point for various approximations and solutions for
special cases. One possibility is to solve (\ref{rekur_p0}--\ref{rekur_pn})
using matrix continued fractions (see e.g. \cite{Risken}).
The continued fraction has to be truncated, it can then be
evaluated numerically. Another possibility is to solve 
(\ref{rekur_p0}--\ref{rekur_pn})
for small $\tau$ perturbatively. This has been done to first order
in \cite{Mi2}, and higher orders can be calculated straight forward.
We give the results up to second order in the next section.

In the special case where the potential $V(x)$ is piecewise linear,
the force $f(x)$ is piecewise constant and (\ref{rekur_p0}--\ref{rekur_pn})
become linear equations with constant coefficients that can be solved
explicitly. The remaining task is to satisfy the continuity conditions
for $p_n(x)$. For the simplest case, a sawtooth potential,
this has been explained in detail in \cite{Mi1,Mi2}. A generalization
to other piecewise linear potentials is straight forward. 
In Section \ref{results} we will present some results for a potential
with three and four pieces. 
\Section{$\tau$--expansion} \label{tauexpansion}
As already discussed in \cite{Mi2}, it is useful to construct
a $\tau$--expansion for constant $\gamma_{n,m}$ in the case of a 
fluctuating potential. When studying a potential driven
by a noisy force one often uses a $\tau$--expansion for
fixed $D=\gamma_{0,1}^2/\tau$. In the case of a fluctuating
potential, however, this would imply arbitrary large potential 
fluctuations, which is clearly unphysical. 

The $\tau$--expansion can be obtained using
standard perturbation theory for linear operators \cite{Kato}.
But it is also possible to start directly from the recursion relations
(\ref{rekur_p0}--\ref{rekur_pn}). To obtain a $\tau$--expansion for 
constant $\gamma_{n,m}$, one uses 
\begin{equation}
  \label{p_tauexpansion}
  p_0(x)=p_{00}(x)+p_{01}(x)\tau+p_{02}(x)\tau^2+O(\tau^3),
\end{equation}
for $p_0(x)$ and similarly
\begin{equation}
  \label{J_tauexpansion}
  J=J_0+J_1\tau+J_2\tau^2+O(\tau^3).
\end{equation}
This ansatz implies that $p_n(x)=O(\tau^n)$. Thus the lowest order
terms can be obtained from (\ref{rekur_p0}), which yields
\begin{equation}
  \label{p00}
  p_{00}(x)=Ce^{\textstyle -\frac{\gamma_{0,0}V(x)}{T}}
\end{equation}
and $J_0=0$. $C$ is fixed using the normalization of $p_{00}(x)$.
In the next order, the term containing $p_1(x)$
in (\ref{rekur_p0}) becomes important. The equation can
now be solved by means of a variation of the constant,
\begin{equation}
  \label{p01}
  p_{01}(x)=C^{(1)}(x)e^{\textstyle -\frac{\gamma_{0,0}V(x)}{T}}
\end{equation}
For $C^{(1)}(x)$ we obtain
\begin{equation}
  \label{C1}
  C^{(1)}(x)=-\frac{1}{T}\int_0^x\,\left(J_1+\gamma_{0,1}^2f(y)
    \left\{\frac{d}{dy}f(y)p_{00}(y)\right\}
  \right)\,e^{\textstyle \frac{\gamma_{0,0}V(y)}{T}}\,dy
  +C_0^{(1)}.
\end{equation}
The constant $C_0^{(1)}$ can be fixed through $\int_0^Lp_{01}(x)dx=0$.
For $J_1$ one obtains using $C^{(1)}(0)=C^{(1)}(L)$
\begin{equation}
  \label{J1}
  J_1=-\frac{\gamma_{0,1}^2}{T}\gamma_{0,0}
  \frac{\int_0^L\, f(x)^3\,dx}
  {\int_0^L\,e^{\textstyle -\frac{\gamma_{0,0}V(x)}{T}}\,dx\,
    \int_0^L\,e^{\textstyle \frac{\gamma_{0,0}V(x)}{T}}\,dx}.
\end{equation}
The sign of $J_1$ depends only on the sign of $\gamma_{0,0}f(x)$.
This means that to first order in $\tau$ one never has a current
reversal. Later in the discussion of our numerical results we
will see that the current changes its sign as a function of
$\tau$. Therefore it is interesting to see whether this behaviour
can be obtained within the $\tau$--expansion. 
The calculations are straight forward, for additional details we refer
to \cite{Mi2}. The final result for the second term $J_2$ is
\begin{eqnarray}
  \label{J2}
  J_2&=&C\tilde{C}\gamma_{0,1}^2\Big\{(5\gamma_{0,0}-\gamma_{1,1})
  \int_0^L \,f'(x)^2f(x)\,dx
  +\frac{\gamma_{0,0}^2(\gamma_{1,1}-\gamma_{0,0})}{T^2}
  \int_0^L\,f(x)^5\,dx\Big\}
  \nonumber\\
  &&+\tilde{C}\gamma_{0,1}^2\Big\{\int_0^L\, C^{(1)}(x)f(x)f'(x)\,dx
  -\frac{\gamma_{0,0}}{T}\int_0^L\,f^3(x)C^{(1)}(x)\,dx \Big\}, 
\end{eqnarray}
with $\tilde{C}=\big[\int_0^L\,\exp(\gamma_{0,0}V(x)/T)\,dx\big]^{-1}$. 
Depending
on $\gamma_{1,1}$ and on $f(x)$ this expression may be positive
or negative. This shows that for sufficiently large $\tau$
one may have a current reversal. 

The $\tau$--expansion presented here has been constructed perturbatively.
The $n$-th order is well defined if the $n$-th derivative of the potential
$V(x)$ exists. Thus the $\tau$--expansion is well defined only for 
analytic potentials, otherwise it breaks down. This happens if
we take a piecewise linear potential. In that case already $J_2$ is
not defined. A similar observation is made in
for a fluctuating force. In \cite{Kohler} it was shown how
one can obtain an asymptotic $\tau$--expansion for that model.
In general one obtains an expansion in $\sqrt{\tau}$. 
But in contrast to the model with a fluctuating force, the first
term $J_1$ in the perturbative $\tau$--expansion calculated above
is always correct in the present case.

\Section{Exact numerical results} \label{results}

As already mentioned above, the recursion relations become
differential equations with constant coefficients if the
force is piecewise constant. Let us divide the interval
$I=[0,L]$ into a set of disjoint intervals $I_k$ and let
us assume that $f(x)=f_k$ if $x\in I_k$. An appropriate
ansatz for $p_n(x)$ is 
\begin{eqnarray}
  \label{ansatz_p0}
  p_0(x)&=& 
  \sum_{r}c_{r,k}a_{0,k}^{(r)}\alpha_k^{(r)}e^{\scriptstyle \alpha_k^{(r)}x}
  +b_{0,k}\mbox{ if  }x \in I_k,
  \\
  \label{ansatz_pn}
  p_n(x)&=&
  \sum_{r}c_{r,k}a_{n,k}^{(r)}e^{\scriptstyle \alpha_k^{(r)}x}
  +b_{n,k} \mbox{ if  }n\ge 1,\,x \in I_k.
\end{eqnarray}
Inserting these expression in the recursion relations one obtains
a generalized eigenvalue problem for the coefficients $a_{n,k}^{(r)}$
and $\alpha_k^{(r)}$,
\begin{equation}
  \label{eigenvalueproblem}
  {\bf A}_k \vec{a}_k=\alpha_k{\bf B}_k\vec{a}_k+T\alpha_k^2\vec{a}_k
\end{equation}
\begin{equation}
  \label{matrixA}
  {\bf A}_k=\left(\begin{array}{*{4}{c@{\hspace{.5ex}}}c}
      0 &  &  &  & \\
      & \lambda_1 & & \mbox{\Large $0$} & \\
      &  & \lambda_2 & & \\
      &\mbox{\Large $0$} & &\ddots & \\
    \end{array} \right),
  \qquad\qquad 
  \vec{a}_k=\left(\begin{array}{c} a_{0,k}\\a_{1,k}\\\vdots\\\end{array}\right)
\end{equation}
\begin{equation}
  \label{matrixB}
  {\bf B}_k=\left(\begin{array}{*{3}{c@{\hspace{.5ex}}}*{2}{c@{\hspace{1ex}}}c}
      \gamma_{0,0}f_k & \gamma_{0,1}f_k & & & & \\
      \gamma_{0,1}f_k & \gamma_{1,1}f_k & \gamma_{1,2f_k} & &
      \mbox{\Large $0$} & \\
      &\gamma_{1,2}f_k  & \ddots &\ddots & & \\
      \mbox{\Large $0$}&  &\ddots &\ddots &\ddots & \\
\end{array} \right).
\end{equation}
For details of the derivation and for an analytical solution of 
this problem for some special noise processes, we refer to \cite{Mi2}.
In general, one can always truncate the matrices ${\bf A}_k$
and ${\bf B}_k$ at some large value of $n$ and solve the eigenvalue
problem numerically. In the following we will discuss results for
sums of dichotomous processes. In this case it is possible to solve 
the eigenvalue problem analytically.
The remaining task is to calculate the coefficients $c_{r,k}$ and the 
current $J$. These quantities can be computed using the continuity
of $p_n(x)$ for $n\ge 0$ and $p_n'(x)$ for $n\ge 1$ at the points where the
force jumps from one value to another. These continuity conditions together
with the normalization of $p_0(x)$ yield a set of linear equations for
the unknown coefficients $c_{r,k}$ in the ansatz
(\ref{ansatz_p0}--\ref{ansatz_pn}) and the 
current $J$. The current can finally be expressed as a ratio of
two determinants. A detailed description of this procedure has
been given in \cite{Mi2} for a sawtooth potential, i.e. for the
case where $f(x)$ takes two values. For a single dichotomous process
the current has a fixed sign, but already for a sum of two or
more dichotomous processes the current may change its sign as
a function of the parameters, e.g. of $\tau$. The main motivation
of the present work was to study the dependence of the current 
as a function of the potential. The simplest case is that of a force
$f(x)$ that takes three instead of two different values, i.e. 
\begin{equation}
  \label{f3}
  f(x)=\left\{ \begin{array}{*{1}{c@{\hspace{1ex}}}c}
      f_1&\qquad\mbox{if } x \in I_1=[0,L_1) \\
      f_2&\qquad\mbox{if } x \in I_2=[L_1,L_2)  \\
      f_3&\qquad\mbox{if } x \in I_3=[L_2,L) 
    \end{array}\right. . 
\end{equation}
The parameters cannot be chosen independently. Using
$L_0\!=\!0$, $L_3\!=\!L$ and introducing the interval lengths 
$\Delta_k:=L_k-L_{k-1}$, we have $\sum_{k=1}^3f_k\Delta_k\!=\!0$
due to the periodicity of $V(x)$. When the potential is close to
a simple sawtooth potential, for example if $f_1$ and $f_2$ are negative,
$f_3$ is positive and $\Delta_3$ is smaller than $\Delta_1+\Delta_2$ one 
observes a behaviour that is similar to what has been obtained in \cite{Mi2}.
But if $\Delta_1+\Delta_2$ is close to $\Delta_3$ new phenomena 
can be observed. 
If $\Delta_1+\Delta_2=\Delta_3$ and $f_1=f_2$, the potential has 
inversion symmetry and consequently $J=0$. In the following our goal will be
to study what happens if one has a small deviation of the inversion symmetric
case. For instance one can take a situation where $\Delta_1$ is much smaller
than $\Delta_2$ and $|f_1|$ is much larger than $|f_2|$. A typical
form of the potential is shown in Figure 1. 
\begin{figure}[ht]
\vspace{3cm}
\leavevmode
\centering
\epsfig{file=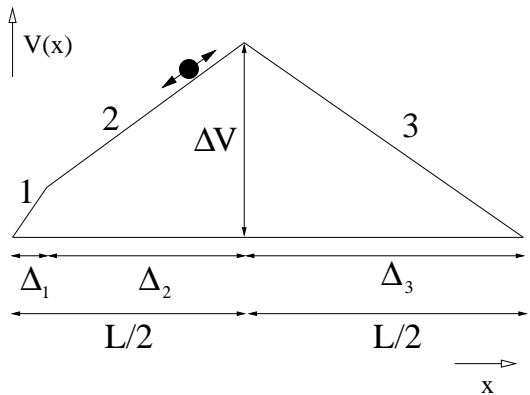,scale=.35,angle=0} 
\caption{A slight symmetry broken three piece linear potential.}
\end{figure}
We begin with the discussion of 
the behaviour of the system if $z(t)$ is a
simple dichotomous process that takes two values $z_1$ and $z_2$ 
with equal probability. 
We will always assume that $\langle z \rangle=\gamma_{0,0}>0$, 
i.e. $z_1+z_2>0$. Then the first term in the $\tau$--expansion 
yields a positive current. It turns out
that, depending on the special choice of the parameters, one obtains 
a current reversal even in the case of a dichotomous noise process.
An example is shown in Figure 2. 
\begin{figure}[ht]
\leavevmode
\centering
\epsfig{file=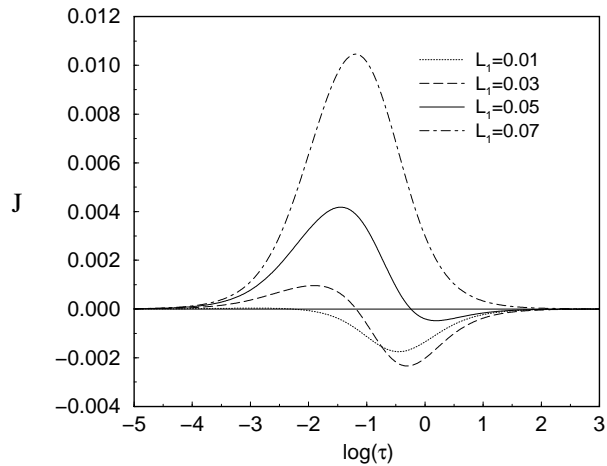,scale=.35,angle=270} 
\caption{The current as a function of $\log(\tau)$ for different
potentials. The parameters of the noise are 
$N=1,\,z_1=0\mbox{ and, }z_2=2$. The temperature is $T=0.04$. 
Each potential has the common parameters $\,L_2=0.5,\,L_3=1,\,f_1=-2,
\,f_3=0.47$ and different values of the interval length $\Delta_1=L_1$
and the slope $f_2$. The corresponding values of $f_2$ are: 
$L_1=0.01\rightarrow f_2\approx-0.439$;
$ L_1=0.03\rightarrow f_2\approx-0.372$;
$L_1=0.05\rightarrow f_2=-0.3$;
$L_1=0.07\rightarrow f_2\approx-0.221$.}
\end{figure}
Here we have $z_1=0$ and
$z_2>0$. Let us first assume that $\tau$ is small. If $z=z_1$, the 
particle is only subject to thermal fluctuations. When the potential 
switches to $z_2$, the probability to find the particle on 
the longest potential slope 3 is the same than to find it on any of 
the two other slopes. For not too large temperatures, the thermal 
perturbation in the equation of motion can be neglected. A simple 
calculation then shows that the time required to reach the minimum 
of the potential along two potential slopes is minimized if the slopes 
are equal. On average the particle moves to the right; as predicted by 
the $\tau$--expansion, a positive current flows. We now consider the 
opposite situation of sufficiently large $\tau$.
Then one has a nearly adiabatic behaviour. But, as shown in \cite{Mi2},
the current vanishes in the adiabatic limit. The first correction
is of order $O(1/\tau)$. Suppose that the temperature is not too large.
If $z=z_2$, the particle will move towards the minimum of the potential
and we can assume that it reaches a nearly static probability distribution
$\propto\exp(-zV(x)/T)$. This distribution is peaked near the minimum,
but due to the asymmetry of the potential the probability to find
the particle on the left hand side of the minimum will be larger than to
find it on the right hand side. When $z$ switches to $z_1=0$, the
particle moves only due to the thermal noise. But since the initial
probability distribution is asymmetric, the probability distribution 
at a finite time will be asymmetric as well. It is simply given by
the convolution of the initial distribution with a Gaussian. Therefore,
when $z$ switches back to $z=z_2$, the probability to reach the next
minimum on the left hand side will be larger than to reach the 
next minimum on the right hand side. This produces a net current 
to the left. Therefore the model shows a current reversal. This can 
be observed in Figure 2. For small values of the interval length 
$\Delta_1=L_1$, the region where the current is positive is very small, 
for larger values of $L_1$ the region where the current is negative 
reduces rapidly.  

It is interesting to study what happens when one varies other
parameters of the system, e.g. the temperature, or what happens for a
more general noise process, which showed already a current reversal in
the case of a sawtooth potential.  In the following we present some
typical results for a potential with the parameters
$f_1=-2,f_3=0.47,L_1=0.03,L_2=0.5$, and $L=1$.  From these values one
obtains $f_2\approx -0.372$ and $\Delta V=|f_3\Delta_3|=0.235$.  We
have chosen these parameters since, as seen in Figure 2, they show the
new current reversal very clearly, whereas for somewhat larger or
smaller values of $L_1$ the regions where the current is negative or 
positive becomes very small.

Let us first discuss the current as a function of the temperature 
for a single dichotomous process. In Figure 3 
\begin{figure}[ht]
\leavevmode
\centering
\epsfig{file=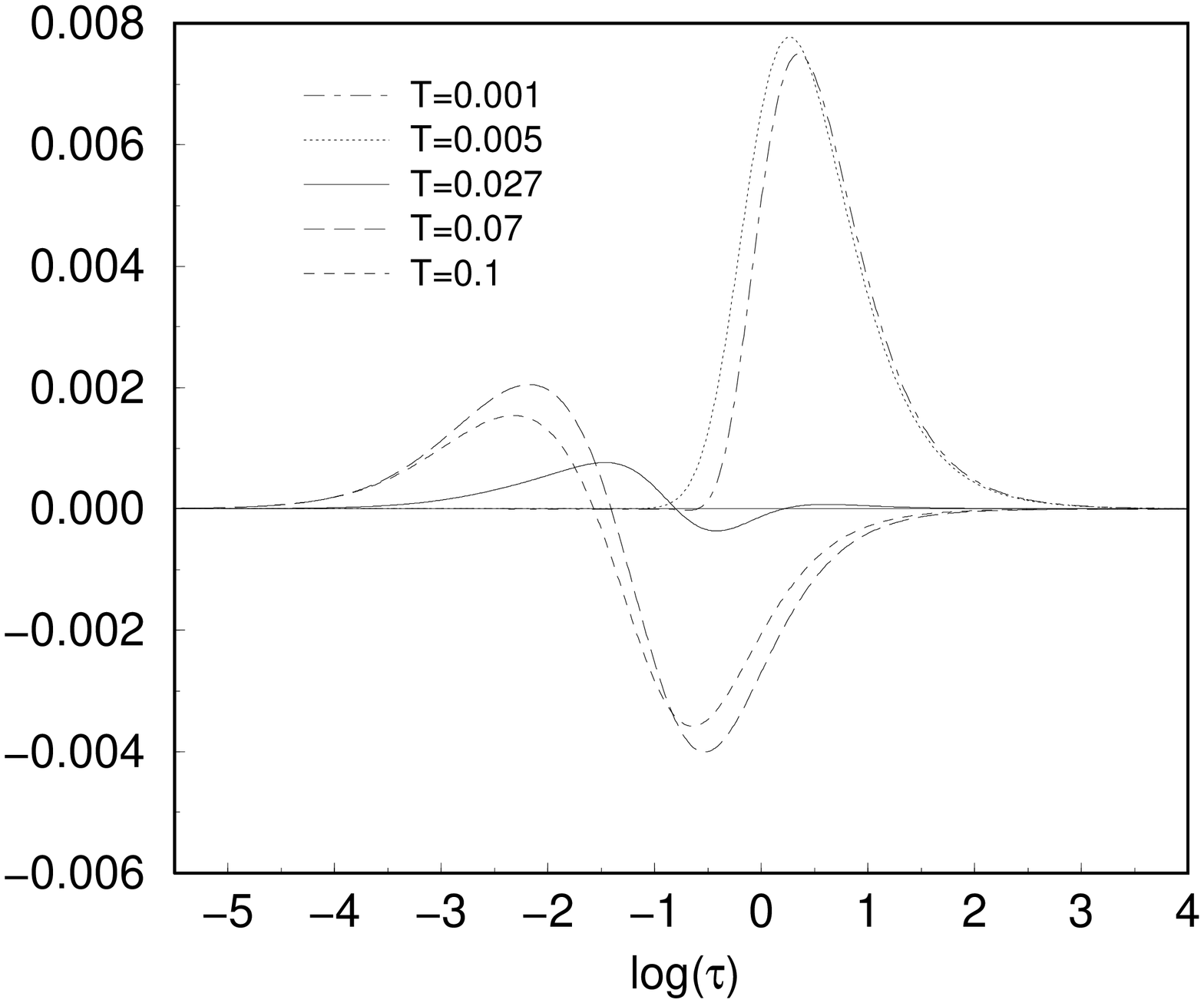,scale=.35,angle=270} 
\caption{The current as a function of $\log(\tau)$ for a symmetric dichotomous 
 process. The two values are $z_1=-0.3$ and $z_2=2.3$.}
\end{figure}
we show results
for the current as a function of $\log(\tau)$ for various temperatures.
The dichotomous process takes the two values 
$z_1=-0.3$ and $z_2=2.3$. One observes that for a small temperature region
around $T=0.027$ the current is positive for small $\tau$, becomes negative
when $\tau$ increases and then becomes positive again. The third
region is strongly enhanced when the temperature is smaller and it vanishes
when the temperature becomes larger. To understand this new effect, we 
first consider the system at temperatures, for which the condition 
$z_1^2\,\mbox{max}\,(f(x))^2\!$
\begin{minipage}[b]{0.4cm}$\vspace{-0.1cm}
\stackrel{<}{\sim}$\end{minipage}
$2T\!\ll z_2^2\, \mbox{min}\,(f(x))^2$ holds. 
Then one basically has the same situation as in the case where one of
the coupling constants is zero and the other positive. When
$z(t)=z_1$, the motion of the particle in the potential is smeared out
by the thermal noise.  If we now slowly turn the temperature down, the
particle will begin to feel the influence of the flat potential.  Let
us consider the nearly adiabatic case at sufficiently low
temperatures.  If $z(t)=z_2$, the particle reaches a nearly stationary
probability distribution. When the potential switches to $z_1$, the
distribution spreads out and the particle diffuses over several
potential valleys.  Since the potential slopes are now much flatter,
the particle has not enough time to reach the stationary distribution
before the potential fluctuates and we can by no means invoke the
stationarity argument used above. Nevertheless, the dynamics is again
determined by the geometry of the potential: Due to the steep slope 1,
the particle surmounts the right slope 3 more easily than the other
two ones (see Fig. 1).  On average, one obtains a positive flow.

Instead of a single dichotomous process we now take a sum
of $N-1$ identical dichotomous processes, each again with
the two values $z_1$ and $z_2$. The static distribution of
$z$ is given by
\begin{equation}
  \label{sd_stat}
  \phi_0(z)=\frac{1}{2^N}\sum_{k=0}^N{N \choose k}\delta(z-(N-k)z_1-kz_2).
\end{equation}
and we have
$z_{1,2}= \frac{\left\langle z\right\rangle}{N}\pm \gamma$, where 
$\gamma^2=\frac{\langle z^2\rangle-\langle z\rangle ^2}{N}$. After 
performing the rescaling  
$\gamma\rightarrow \frac{\gamma}{\sqrt{N}}$, the coefficients 
$\gamma_{n,m}$ are given by
\begin{equation}
  \label{sd_gamma}
  \gamma_{n,n+1}^2=\frac{\left(n+1\right)\left(N-n\right)}{N}\gamma^2,\quad 
  \gamma_{n,n}=\left\langle z\right\rangle. 
\end{equation}
In \cite{Mi2} it was shown that for this class of processes the
eigenvalue problem (\ref{eigenvalueproblem}) can be solved analytically.

The question is now whether the effect observed for a single dichotomous
process can be observed in this case as well, and if perhaps additional
new effects occur. Next we present some results for a sum of
two dichotomous processes. Let us first discuss the behaviour for large
values of $\tau$. As in the case of a single dichotomous process we
have a nearly adiabatic behaviour. But now, $z$ takes three values
$2z_1$, $z_1+z_2$, and $2z_2$. For $z_1<0$, $z_1+z_2>0$ and small
temperature the behaviour should be that of a single dichotomous
process that takes the two values $2z_1$, and $z_1+z_2$. 
 Therefore, the sign of the current depends on the difference of 
$z_1$ and $z_2$, i.e. on $\gamma$. If the condition 
$\left|2z_1\right|<z_1+z_2$ holds, i.e. if  
$\gamma< \sqrt{2}\,\langle z\rangle$, it will be positive and otherwise, 
negative. Figures 4 and 5 confirm this qualitative discussion.
\begin{figure}[p]
\leavevmode
\centering
\epsfig{file=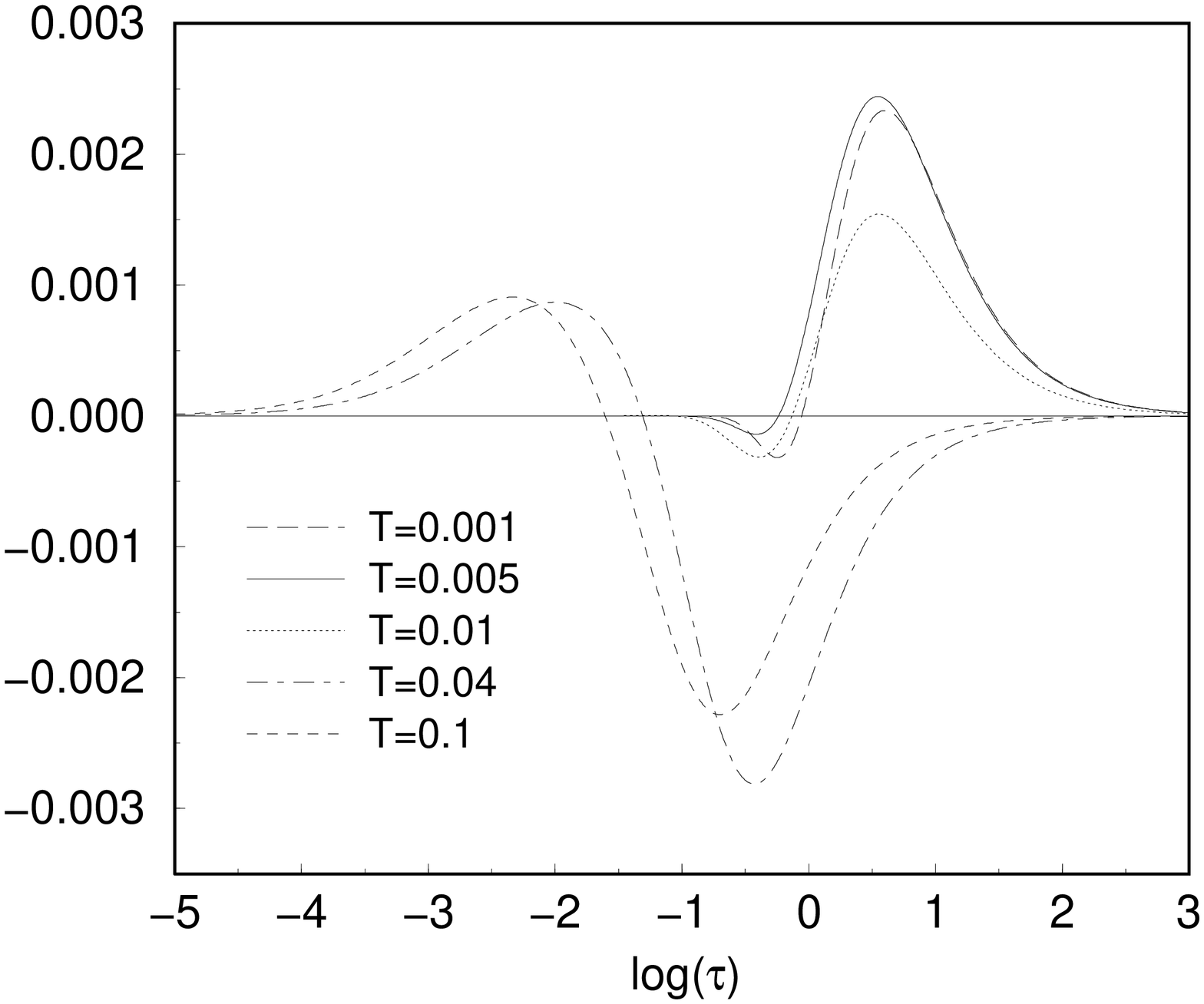,scale=.35,angle=270}
\caption{The current as a function of $\log(\tau)$ for a sum of two 
dichotomous processes at different temperatures. The average of the coupling 
constant and the noise strength are $\langle z\rangle=1$ and $\gamma=1$.} 
\end{figure}
\begin{figure}[p]
\leavevmode
\centering
\epsfig{file=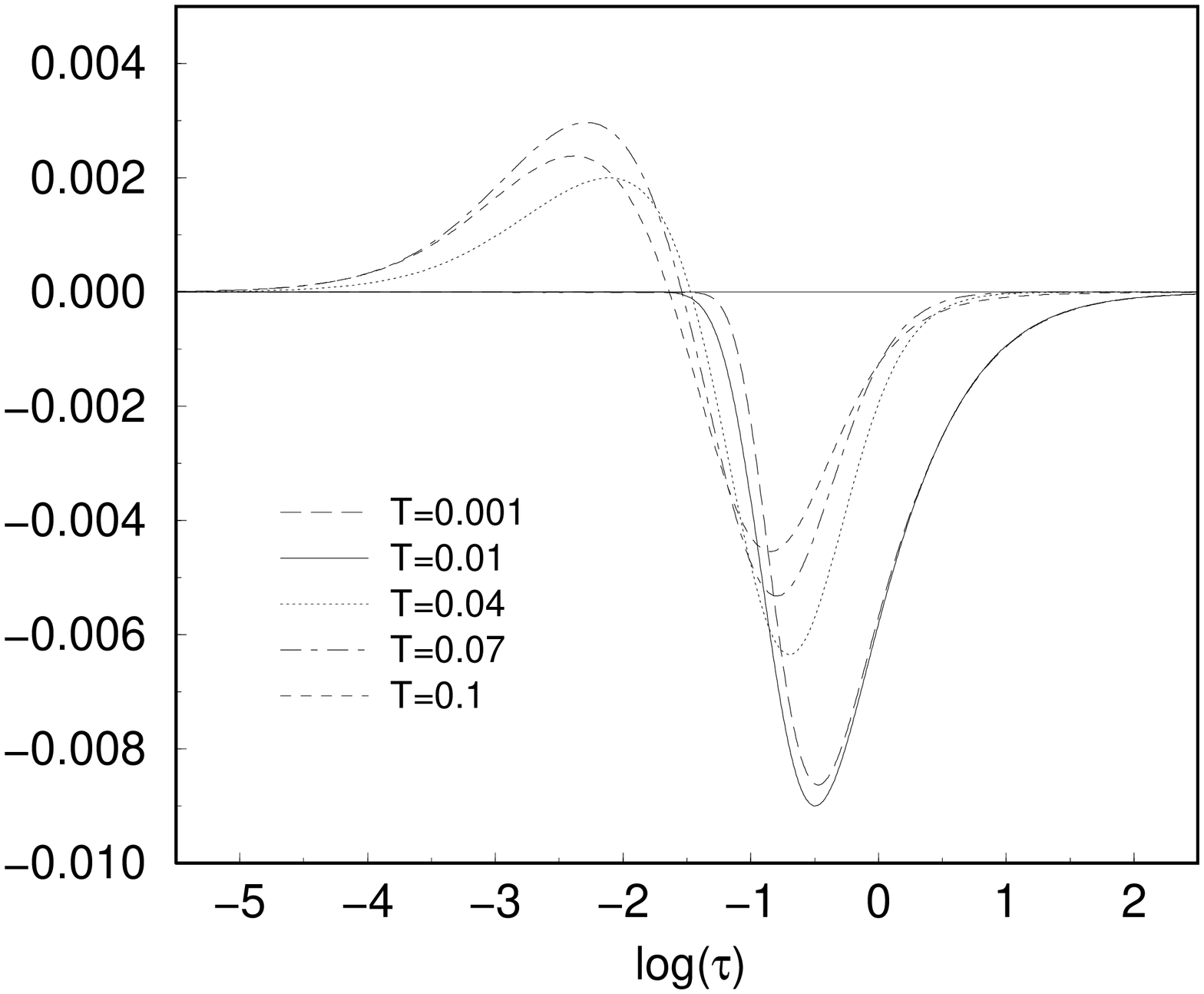,scale=.35,angle=270}
\caption{The current as a function of $\log(\tau)$ for a sum of two
dichotomous processes at different temperatures. The parameters of the 
noise are $\langle z\rangle=1\mbox{ and } \gamma=1.7$.} 
\end{figure}

At higher temperatures, the motion in the potential $2z_1V(x)$ 
is smeared out and the system 
behaves again as in the case of a simple dichotomous process with $z_1=0$ 
and $z_2>0$. From our discussion, we expect only one current reversal for 
all $\gamma> \sqrt{2}\,\langle z\rangle$. However, it turns out that for 
higher values of $\gamma$ and sufficiently large temperatures a weak 
positive current is observed when $\tau$ becomes large (Figure 6). 
\begin{figure}[ht]
\leavevmode
\centering
\epsfig{file=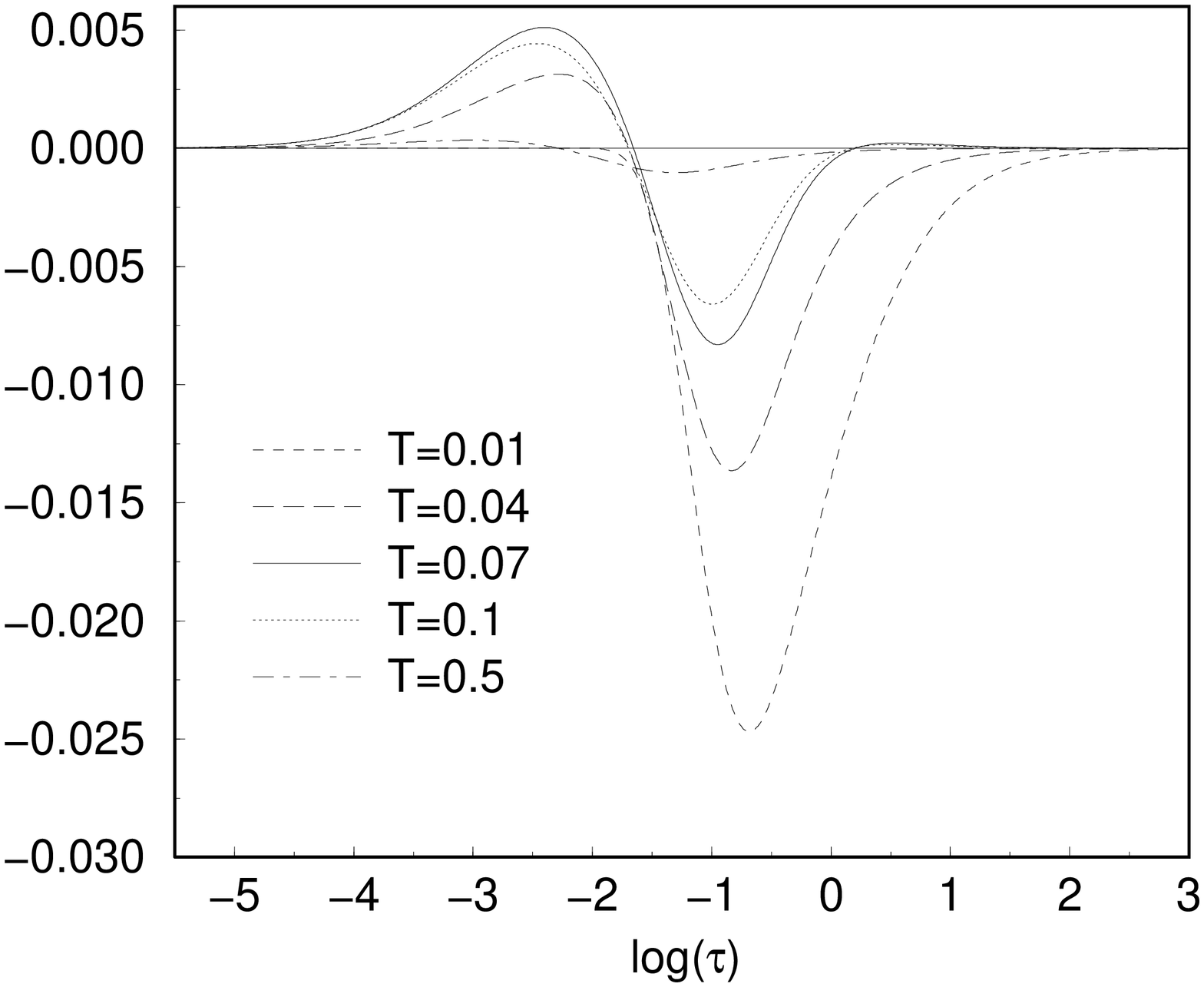,scale=.35,angle=270}
\caption{The current as a function of $\log(\tau)$ for a sum of two dichotomous
processes at different temperatures. Here we have set 
$\langle z\rangle=1 \mbox{ and }\gamma=2.5$. } 
\end{figure}
To 
understand the different behaviour in Figure. 5 and 6, first note that 
for $\gamma=1.7$ the quantities $|2z_1|=1.4$ and $z_1+z_2=1$ have 
a similar numerical value, whereas for $\gamma=2.5$ one has  
$2z_1\approx -3.4$ and $z_1+z_2=1$. 
In the first case the motion for both values $z(t)$ 
is determined mainly by the thermal noise when $T$ rises; but in the second, 
only the motion in the potential that 
corresponds to the value $z(t)=\langle z\rangle$ 
will be smoothed out while the motion in the 
potentials for the other two values  
is not influenced so much by the thermal noise. 
For large $\tau$ and not too high 
temperatures only the values with small $\left| z(t)\right|$ are relevant, 
i.e. $2z_1$ and $z_1+z_2$. Thus the system behaves as in the case of a 
single process where one of the coupling constants is zero and the other 
is negative. This yields a small positive current.

Let us briefly discuss the behaviour of the current for a sum of $N$ 
identical symmetric dichotomous processes. Figure 7 
\begin{figure}[ht]
\leavevmode
\centering
\epsfig{file=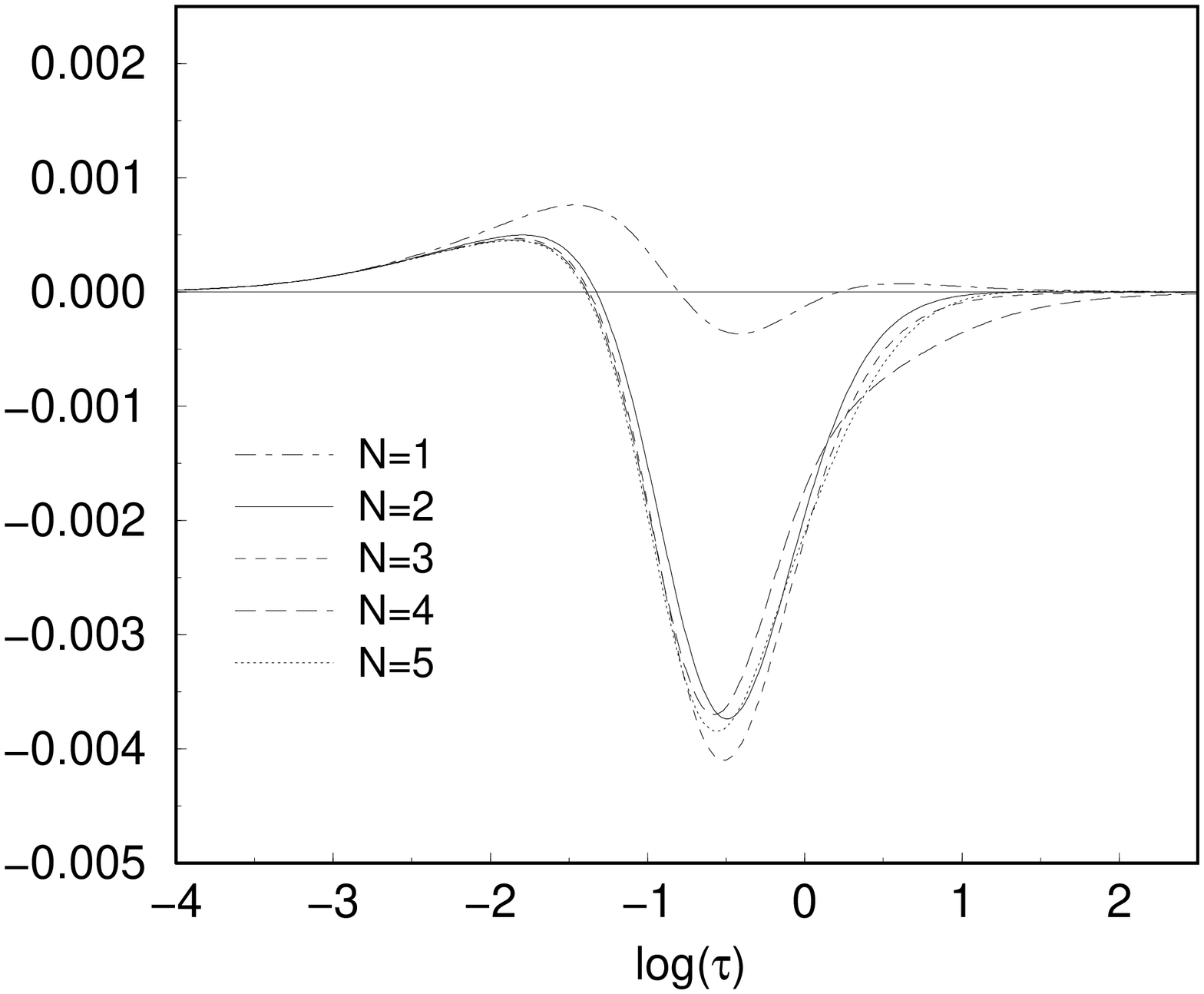,scale=.35,angle=270}
\caption{The current as a function of $\log(\tau)$ for a sum of dichotomous 
processes. $N$ takes the values 1,2,3,4 and 5. The other parameters of the
noise read $\langle z\rangle=1$ and $\gamma=1.3$; the temperature is 
$T=0.027$.} 
\end{figure}
shows some plots of the 
current as a function of $\tau$ for different values of $N$. For small values 
of the correlation time ($< 10^{-2}$), all curves merge into a single curve.
This is a generic feature, which holds also for general potentials $V(x)$. 
The reason is that, according to (\ref{J1}) and (\ref{J2}), 
the coefficients $J_1$ and 
$J_2$ of the $\tau$--expansion are the same for all $N$, since the quantities 
$\gamma_{0,0}$, $\gamma_{0,1}$ and $\gamma_{1,1}$ do not depend on $N$. 
In contrast, the behaviour of each curve is different near 
the adiabatic limit; for $N>1$, the effect responsible for a 
positive current disappears.

The fact that all plots lie so close to each other 
suggests that they rapidly converge to a single curve for $N\to\infty$. 
In this limit, the sum of dichotomous processes yields a well known 
Gaussian process, the so-called Ornstein--Uhlenbeck process
\cite{Maruyama}. In this 
case the potential, rather than jumping between different states, 
suffers a continuous distortion. One therefore expects that 
some mechanisms that lead to a current reversal cancel. 
Nevertheless, our results show that the current 
changes its sign once, in contrast to the case of a sawtooth potential, 
where the sign of the current is fixed \cite{Mi2}.
This behaviour is also predicted by the $\tau$--expansion, which
leads to a current reversal for potentials of the form
discussed above and an Ornstein--Uhlenbeck process. 

The validity of the expansion 
around the white noise limit can be tested
with the help of the exact numerical results. 
In Figure 8 
\begin{figure}[ht]
\leavevmode
\centering
\epsfig{file=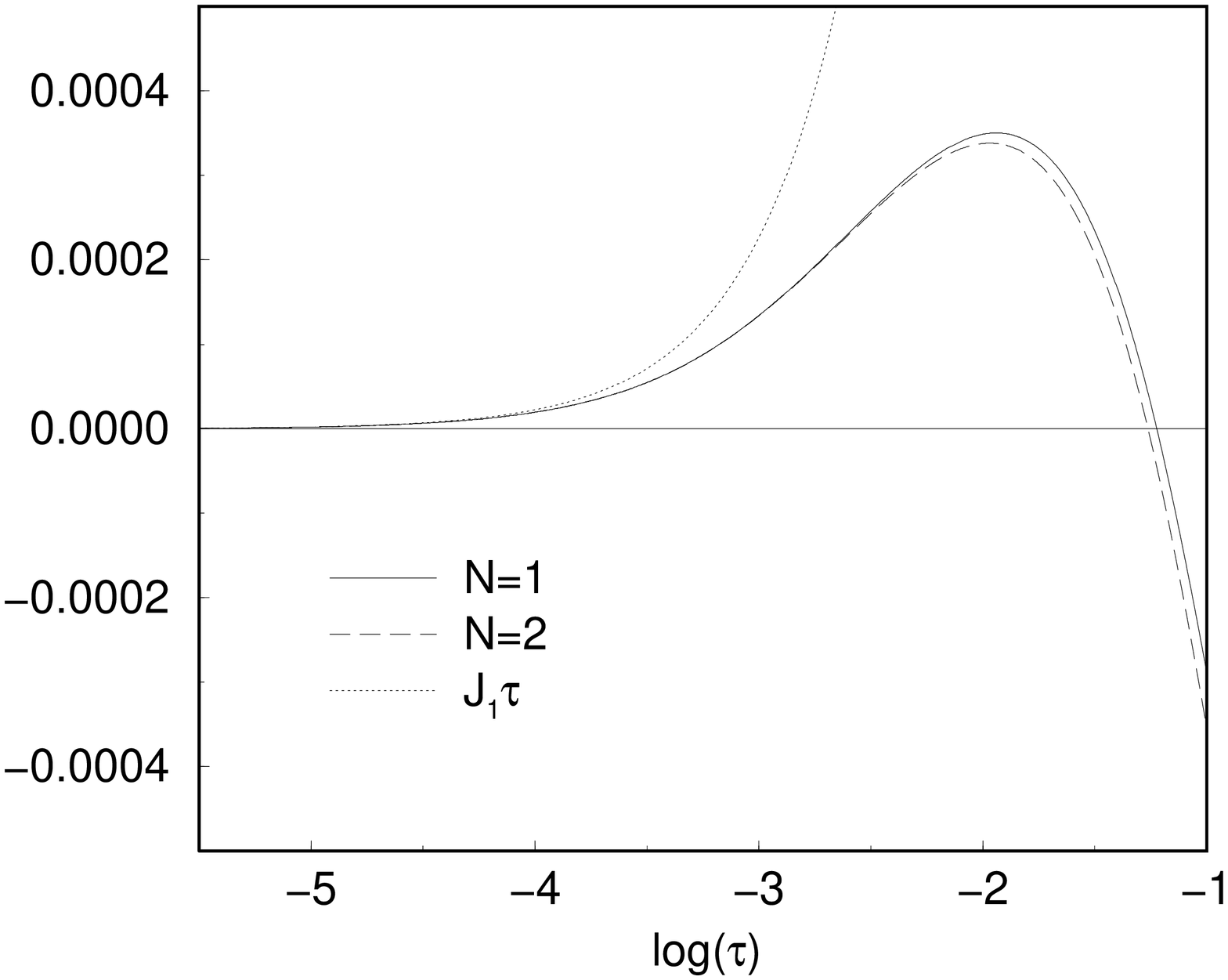,scale=.35,angle=270} 
\caption{Two plots of the current as a function of $\log(\tau)$ computed for 
a simple dichotomous process (solid line) and a sum of two 
processes (dashed line) are shown together with the 
first order term (dotted line) of the $\tau$--expansion. 
Here we have set $\langle z\rangle=1
\mbox{ and }\gamma=0.6$. The temperature is $T=0.04$.} 
\end{figure}
two exact plots of the current for $N=1$ and $N=2$ and the 
first order term of the $\tau$ expansion are 
depicted. The quantitative agreement 
with the exact curves is remarkably good for values of $\tau$ below 
$10^{-4}$. 

To conclude this section let us discuss briefly what happens for more
complicated potentials. It is straight forward to do the same calculations
as above for e.g. a piecewise linear potential with four (or more) pieces.
We have done several calculations for a potential with four pieces.
The results for small and large values of $\tau$ are similar to the 
results reported for the potential with three pieces. For small $\tau$
a current reversal may occur and can be predicted by the $\tau$--expansion.
For large $\tau$, in the nearly adiabatic limit, the current is positive
or negative, depending on the temperature and the noise parameters.
The mechanism we discussed in that case for the potential with three pieces
can be carried over to the more general case. For intermediate values
of $\tau$, the current shows a more complicated behaviour. Depending
on $T$ and $\gamma$, one observes additional minima and maxima for
$J$ as a function of $\tau$. On the other hand, we never observed more
than two current reversals for $J$ as a function of $\tau$. 

\Section{Conclusions}
The main result of this paper is that a new kind of current reversal
occurs for a Brownian particle in a fluctuating potential, if the
potential is not a simple sawtooth potential. For a saw-tooth
potential a current reversal has only been observed for more
complicated noise processes, but not for a simple dichotomous process
or for the Ornstein--Uhlenbeck process \cite{Ast1,Prost,Mi2}.  In
a more complicated potential a current reversal even occurs for the
dichotomous process and the Ornstein--Uhlenbeck process.  It can even
be predicted by a simple $\tau$--expansion carried out to second
order.  For a general potential we have derived the first two
non-vanishing terms of the expansion around the white noise limit and
we have shown that, depending on the noise parameters and the shape of
the potential, they may have different signs, thus leading to a
current reversal.

The stationary current depends strongly both on the statistics of the 
coloured fluctuations and the details of the potential. Depending on
the range of $\tau$ and $T$, the current has a different direction and
magnitude. 

We have focused on the special case of a piecewise linear potential.
Our numerical results show that small deviations of the inversion
symmetric case change the qualitative behaviour of the system dramatically.
If one takes for instance a slight asymmetric three piece linear 
potential, the current may change its sign as a function of $\tau$ 
more than once, even in the case of a simple dichotomous process.  
The number of current reversals depends on the relative sign of the 
coupling constants $z_1$ and $z_2$: If both are positive or one is 
positive and the other one is zero, only one current reversal is 
observed, whereas in the case $z_1<0$ and $z_1+z_2>0$ a new low 
temperature effect takes place and provokes an additional current 
reversal for large $\tau$. In this situation, the induced current 
tends to a finite value when $T\to 0$.

Depending on the time scale of the fluctuations, the asymmetry of the
potential acts on the particle in a different way: In the nearly
adiabatic limit, the thermally driven particle surmounts one of the
adjacent potential barriers more easily than the other one, whereas 
for small $\tau$ (and not too high temperatures), the different drift 
times down the three potential slopes favour one direction of movement. 
Both effects are purely dynamical, it is not possible to explain the
behaviour using transition rates calculated in the adiabatic limit.

For a sum of two dichotomous processes, the system may be described by
an effective dichotomous process if $\tau$ is sufficiently large. The
behaviour depends on the sign of the average of the coupling constant
for the effective process, i.e. on $\langle z\rangle$ and $\gamma$.
When $T$ rises, some values of $z(t)$ will be washed out by the
thermal noise. Therefore, the choice of the coupling constants of the
effective process depends on temperature.
  
In the general case of a sum of $N$ dichotomous processes, we have
seen that the current does not depend on the noise details for
sufficiently small $\tau$.  Whereas in the case of a sawtooth
potential the current changes sign only if $N$ is odd \cite{Mi2}, for
a proper form of the three piece potential one observes at least one
current reversal for all values of $N$.

Since our numerical results have been obtained only
for a piecewise linear potential, 
we would like to emphasize that our results hold
for general potentials as well. As mentioned above the current
reversal in the case of a dichotomous process or in the
case of the Ornstein--Uhlenbeck process can be predicted by
a second order $\tau$--expansion for any given potential.
Furthermore, the argument that for large $\tau$ the behaviour
of the system can be described by an effective dichotomous process,
is true for a general potential as well.
 
Let us discuss the relevance of our results for the motion of
molecular motors. In \cite{Mi2} it was argued that the basic features
of the movement can be described by a sum of $N-1$ dichotomous
processes, where $N\ge 3$ is the number of conformational changes of
the motor protein. Even if one takes a simple sawtooth ratchet as
interaction potential, this model leads to a current reversal for a
proper choice of the parameters.  According to Astumian and Bier this
means that, after redimensionalizing  the equation of motion 
(\ref{lan}), two proteins with a slight different geometry ---and 
thus different friction coefficients--- may drift in opposite 
directions \cite{Ast2}.

On the other hand, our results emphasize that rate and direction of
the movement are very sensitive to the details of the potential.  They
suggest that slight modifications of the binding energy profile which
the motor protein feels when it walks on the periodic polymer
may change its direction of movement.  Even though the polymer tracks
are much more symmetry broken than the potentials that we have
investigated, it should be clear that a sawtooth ratchet is a too
rough approximation for the actual interaction potential. Due
to our results, one would expect that a small deviation of the simple
sawtooth ratchet has a large effect on the current, especially in the
region of intermediate $\tau$. Therefore one should be careful:
A quantitative agreement between calculated currents and experimental
findings may be accidental. But the qualitative
aspect of the model may be correct. 

Clearly, the model is too simple to provide a realistic description
of the situation in a cell.
The kinetics of the ATP hydrolysis in the cell is very sensitive to 
temperature changes of the system. For this reason, $\tau$ and $T$ 
cannot be chosen independently. Another serious difficulty is that 
the forces exerted by the protein on the track during the ATP hydrolysis
change not only the barrier heights but also the shape of the interaction 
potential. This conformational flexibility is not taken into account in
the model, where the interval lengths $\Delta_i$ and the relative heights
of the potential teeth are time independent. But a more realistic
model will contain many free parameters and, as we have shown, the
results may depend strongly on each of these parameters. Therefore
a more detailed experimental knowledge of the system is required.
                                                                              
\clearpage


\end{document}